\documentclass[aps,prl,showpacs,noshowkeys,amsmath,amssymb,amsfonts,reprint]{revtex4-1}
\usepackage{graphicx}
\usepackage{dcolumn}
\usepackage{bm}
\begin{document}
\title{Colloidal Microworms Propelling via a Cooperative Hydrodynamic Conveyor-Belt}
\author{Fernando Martinez-Pedrero$^1$}
\author{Antonio Ortiz-Ambriz$^1$}
\author{Ignacio Pagonabarraga $^{2,3}$}
\author{Pietro Tierno$^{1,3}$}
\email{ptierno@ub.edu}
\affiliation{
$^1$Estructura i Constituents de la Mat\`eria, Universitat de Barcelona, 08028 Barcelona, Spain.\\
$^2$Departament de F\'isica Fonamental, Universitat de Barcelona, 08028 Barcelona, Spain.\\
$^3$Institut de Nanoci\`encia i Nanotecnologia, IN$^2$UB, Universitat de Barcelona, Barcelona, Spain.}
\date{\today}
\begin{abstract}
We study propulsion arising from
microscopic colloidal rotors
dynamically assembled and driven in a viscous fluid
upon application of an elliptically polarized rotating magnetic field.
Close to a confining plate,
the motion of this self-assembled microscopic worm
results from the cooperative flow generated by the spinning particles
which act as a hydrodynamic "conveyor-belt".
Chains of rotors
propel faster than individual ones, until reaching
a saturation speed at distances where induced-flow additivity vanishes.
By combining experiments
and theoretical arguments,
we elucidate the mechanism of motion
and fully characterize the
propulsion speed in terms
of the field parameters.
\end{abstract}
\maketitle
Propulsion in viscous fluids plays a key role in many 
different contexts of biology, physics, and chemistry. 
From  a fundamental perspective,
the transport of microscopic objects in a liquid medium
poses the appealing challenge to find an adequate swimming
strategy due to the negligible role of inertial force compared
to viscous one.
At low Reynolds number the Navier-Stokes equations become time
reversible~\cite{Happ}, and any strategy based on  reciprocal
motion, {\sl i.e.} a motion composed by symmetric backward and forward displacements,
will fail to produce net propulsion~\cite{Pur97}.\\
Facing this challenge, the last few years have witnessed
the theoretical propositions
of several suitable geometries and procedures
to propel micromachines
in viscous fluids~\cite{Sto96,Cam99,Naj04,Lau04,Ale08,Dow09,Leo12}.
Parallel advances in miniaturization
have led to the generation of new classes of
chemically powered~\cite{Pax04,How07}
or externally actuated~\cite{Dre05,Sne09,Vol11,Bri13}
prototypes
with exciting applications in emerging fields
such as microsurgery~\cite{Nel10,Pey13} or
lab-on-a-chip technology~\cite{San11,Wan12}.\\
In contrast to reaction driven microswimmers,
actuated magnetic micropropellers
do not present the autonomous behaviour which distinguishes
force- and torque-free motion of biological organisms~\cite{Lau09},
but instead avoid
efficiency reduction due to fuel shortage or directional randomization.
To date, three main approaches
have been developed to
transport microscopic particles in a viscous fluid using
uniform magnetic fields~\cite{Fis11}, namely
by actuating flexible magnetic tails~\cite{Dre05,Pak11},
by rotating helical shaped structures~\cite{Zha09,Fis09},
or by using the close proximity to
a bounding wall~\cite{Tie08,Zha10}.
In the latter case,
it is well established that
in the Stokes regime the rotational motion of a
body close to a surface can be rectified into net translation due to the
hydrodynamic interaction with the boundary~\cite{Gol67}.
Surface rotors are optimal to work in confined geometries
such as microfluidic channels
or biological networks characterized by narrow pores.\\
In this Letter
we show that an ensemble of rotors dynamically self-assembled
via attractive dipolar interactions
can be propelled by a
hydrodynamic "conveyor-belt" effect
generated by the cooperative
flow of the spinning particles
close to a surface.
Recent theoretical works~\cite{Leo10,Fil12,Ngu14}
have addressed the rich dynamics of
active rotors lying and spinning in the same plane.
Here we rotate our particles perpendicular to the bounding plane
and use real time/space experiments
to elucidate the mechanism
underlying the collective propulsion.\\
We use monodisperse paramagnetic colloids of
radius $a=1.4\mu m$, composed by a polymer matrix evenly doped with
nanoscale superparamagnetic iron oxide
grains (Dynabeads M-270, Invitrogen).
The particles are diluted in Millipore
water and deposited
above a glass substrate where they sediment due to
density mismatch.
After sedimentation,
each particle floats at a distance $h$
from the plate
due to the balance between gravity and electrostatic
interactions with the negative charged substrate.
The particles display small Brownian motion
above the substrate ($(x,y)$ plane),
with a diffusion coefficient $D=0.14\mu m^2 s^{-1}$,
and negligible out-of-plane fluctuations.\\
As shown in the schematic of Fig.1(a),
we have assembled the magnetic particles
into translating chains by applying
a rotating magnetic field elliptically polarized
in the $(x,z)$ plane with angular frequency $\omega$,
${\bm H}(t) \equiv H_x \cos{(\omega t)}{\bm e}_x- H_z \sin{(\omega t)}{\bm e}_z$.
The amplitude $H_0$ and  ellipticity $\beta \in [-1,1]$ of the applied field
are defined as $H_0=\sqrt{(H_x^2+H_z^2)/2}$ and $\beta=(H_x^2-H_z^2)/(H_x^2+H_z^2)$, respectively.
The applied modulation induces a finite magnetic torque
on the particles, ${\bm T}_m=\mu_0 \langle {\bm m} \times {\bm H} \rangle$,
which sets them in rotation at an angular speed ${\bm \Omega}$
close to the plane.
Here $\mu_0 = 4 \pi \cdot 10^{-7}  H \, m^{-1}$, ${\bm m}$ is the particle magnetic moment
and $\langle ... \rangle$ denotes a time average.
For high driving frequencies, $\omega \gtrsim 50 rad\, s^{-1}$,
the magnetic torque arises due to a finite
internal relaxation time of the particle magnetization~\cite{Tie07,Jan09,Ceb11}, $t_{r}$,
which in our case is $t_{r}\sim 10^{-4}s$.
Upon balancing ${\bm T}_m$ with
the viscous torque
arising from the rotation in the medium ${\bm T}_v=-8\pi \eta {\bm \Omega}a^3$
the average rotational speed reads
$\langle \Omega \rangle = H_0^2\sqrt{1-\beta^2}\chi t_{r} \omega /6\eta (1+ t_{r}^2 \omega^2)$~\cite{EPAPS}.
Here $\chi=0.4$
is the magnetic susceptibility under static field
and $\eta=10^{-3} Pa \cdot s$ denotes the dynamic viscosity of water.
Individual rotors translate at a speed
$\langle {\bf v}_0 \rangle \sim \langle \Omega \rangle a {\bm e}_x$
due to the close proximity of the surface
which breaks the symmetry.
\begin{figure}[t]
\begin{center}
\includegraphics[width=\columnwidth,keepaspectratio]{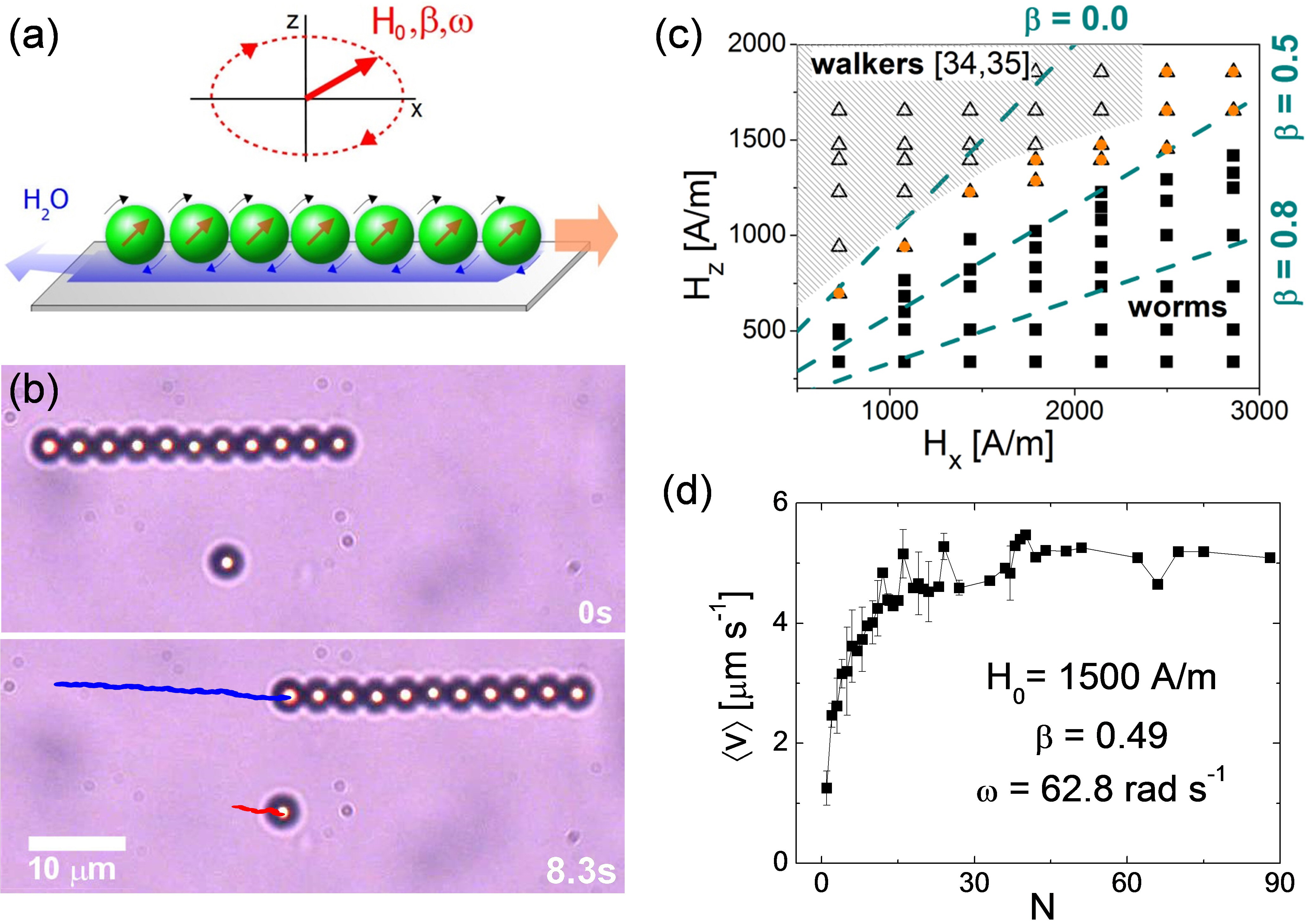}
\caption{(Color online) (a) Schematic of an ensemble of
paramagnetic colloids rotating close to a glass plate and subjected
to an elliptically polarized rotating field.
(b) Microscope images
showing the motion of a chain (blue line) and of an individual
rotor  (red line) subjected to a rotating field
with $H_0 = 1500A \, m^{-1}$ ,
$\beta = 0.49$, and $\omega = 62.8 \, rad s^{-1}$ at $t = 0$ s (top) and
$8.3$ s later (bottom). (c) Regions in the ($H_x,H_z$)
plane where
colloidal "walkers"~\cite{Mor08,Sin10} (green), 
translating worms (white) or both types of dynamics (orange points) are observed.
 ($\omega=62.8\, rad\, s^{-1}$)
(d) Average speed of the colloidal worms $\langle {\mathrm v} \rangle$
{\it vs} number of particles $N$.}
\label{fig_1}
\end{center}
\end{figure}
A pair of particles, $(i,j)$,  a distance
${\bm r}_{ij}$ away, experience a magnetic dipolar interaction of energy,
$U_{d}=(\mu_0/ 4\pi)[ ({\bm m}_i \cdot {\bm m}_j) /r_{ij}^3 -  3({\bm m}_i \cdot {\bm r}_{ij})({\bm m}_j\cdot {\bm r}_{ij}) /r_{ij}^5 ] $,
which is maximally attractive (repulsive) for particles with
magnetic moments parallel (normal) to ${\bm r}$.
Performing a time average of the dipolar energy between
two colloids for a rotating magnetic field  in the $(x,z)$ plane ($\beta=0$),
reveals an  effective attractive potential
in this plane $\langle U_d \rangle =-\frac{\mu_0 m^2}{8 \pi (x+z)^3}$,
leading to chaining
along the $x$-direction,
while
being repulsive in the
perpendicular one, $\langle U_d \rangle =\frac{\mu_0 m^2}{4 \pi y^3}$
(similar results holds for $\beta \neq 0$).
Thus the actuating magnetic field forces
the paramagnetic colloids to assemble into elongated chains or "worms",
while it also ensures the rotational motion of the individual units.\\
The dynamics of a colloidal worm
is illustrated by the two
microscopic images in Fig.1(b)
(VideoS1 in~\cite{EPAPS}).
Under the conditions of this experiment (caption Fig.1),
a chain of $N=11$ rotors covers a distance of $\sim 25\mu m$
at an average speed of $\langle {\mathrm v} \rangle = 3.0\mu m s^{-1}$.
In the same pair of images, below the traveling chain, 
one individual rotor is shown covering
a smaller distance of $\sim 5.6\mu m$ at a lower speed of $\langle {\mathrm v}_0 \rangle = 0.6 \, \mu m s^{-1}$,
making evident the enhanced propulsive behaviour
of the colloidal worm.\\
In order to identify  the range of field parameters
where propelling worms are observed, we perform a series
of experiments by varying the two amplitudes of the applied
field, ($H_x$,$H_z$) which correspond to values of
$\beta \in [-0.73,0.97]$ and $H_0 \in [600, 2400]A \, m^{-1}$,
and at a fixed angular frequency
of $\omega = 62.8\, rad\, s^{-1}$. 
The graph in Fig.1(c) shows the existence of two distinct types of dynamics: for low ellipticity ($\beta<0.2$), the chains rotate as a whole, performing a 3D "walking"-like motion as observed in previous works~\cite{Mor08,Sin10}. In contrast, for larger values of $\beta>0.45$, where the out of plane component of the field $H_z\ll H_x$, we observe translating worms. Note that in the intermediate region both types of dynamic states can be found.
At fixed frequency (data with different $\omega$ are included in~\cite{EPAPS}) walkers can be transformed into worms by increasing the field ellipticity. This effect is caused by the decrease in the azimuthal part of the dipolar force $F_z \sim H_z\frac{\partial H_z}{\partial z}$ 
due to the corresponding decrease in $H_z$, and hence the relative vertical displacement of the colloids, 
in favor of the in-plane part ($H_x$) of the rotating field during one oscillation period. 
As a consequence, increasing $\beta$ at constant $\omega$ decreases the oscillations of the phase-lag angle between the main axis of the chain and the rotating field~\cite{EPAPS}.
In the limit case of $\beta=1$ worms still form but the chains do not show a net motion.
As shown in Fig.1(d),
at a fixed ellipticity of $\beta=0.49$,
and amplitude of $H_0=1500A \, m^{-1}$,
the average speed $\langle {\mathrm v} \rangle$ initially increases with the worm length,
and later saturates to a maximum value of $5.3$ $\mu m s^{-1}$.
Such a trend of the velocity in terms of the number of rotors
is observed both by varying the frequency,
\begin{figure*}[t]
\begin{center}
\includegraphics[width=0.97\textwidth]{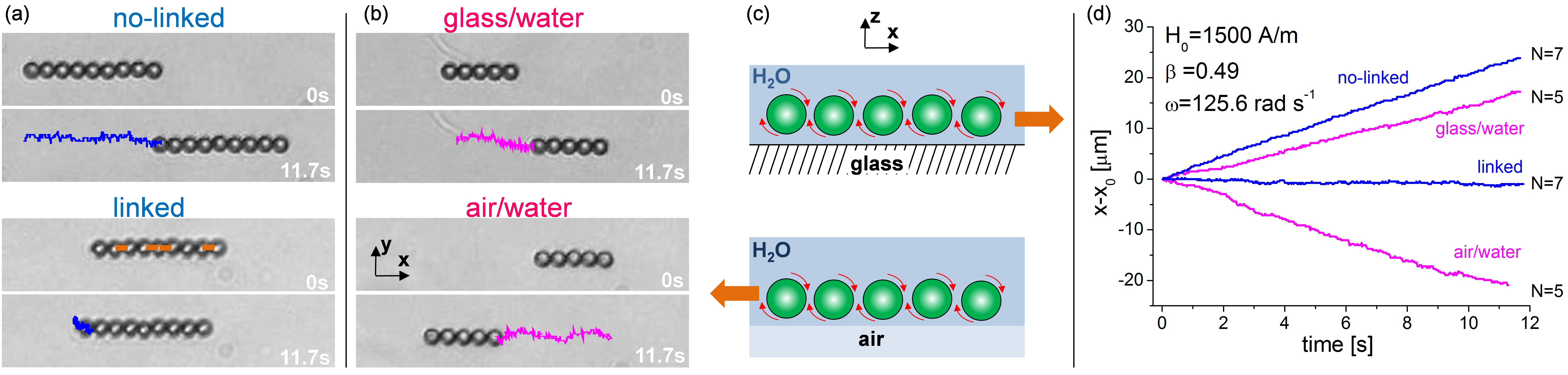}
\caption{(Color online)(a) Sequence of images showing
a worm composed by $N=7$ free rotors (top) or rotors with pairs of
particles permanently linked (orange segments, bottom).
(b) Propelling chain of $N=5$ rotors at glass/water interface
(top) and at air/water interface (bottom). Field parameters for
(a) and (b) are $H_0=1500 A \, m^{-1}$, $\beta=0.49$ and $\omega=125$rad \,s$^{-1}$.
(c) Schematics illustrating the rotors
locations with respect to the interface in (b).
(d) Position ($x-x_0$) versus time for the chains of
particles. Black (blue) line refers to (a), grey (pink) to (b).}
\label{fig_2}
\end{center}
\end{figure*}
or the field amplitude $H_0$.\\
The series of experiments displayed in Fig.2 serve
to  elucidate the mechanism underlying the motion of the colloidal worms.
Fig.2(a) demonstrates that the rotation of the individual
particles within the chain is an essential ingredient
for its net displacement. At the top part of Fig.2(a) a chain of $N=7$
particles is propelled towards the right at
an average speed of $\langle {\mathrm v} \rangle = 2.0\mu m s^{-1}$.
The pair of images at the bottom of Fig.2(a) shows a chain composed
by the same number of particles, but where some of them
have been previously permanently linked by screening the
electrostatic interactions (orange segments)~\cite{Note1}.
We observe no net translational  motion
when subjecting the
linked particles to the same field conditions, 
as shown in VideoS2 in~\cite{EPAPS}.\\
Next, in a separate experiment, we analyze the effect of the
boundary on the propulsion direction and speed of the colloidal worms.
In Fig.2(b) we compare two chains composed by $N=5$ particles
and subjected to the same field conditions, but
propelling close to a glass/water interface (images at the top),
or to a water/air interface (bottom).
The water/air interface was realized by suspending $\sim0.2g$
of surfactant-free water in a specially prepared circular cavity 
of volume of $0.18cm^3$ and 
provided with a
small hole ($2mm$ diameter) at the bottom. Surface tension
is able to hold the deposited droplet
while
evaporation effects are negligible during the experimental
time (see~\cite{EPAPS} for details). 
The paramagnetic colloids sediment approaching
the water/air interface. However as observed previously~\cite{Hel08,Ebe09},
these particles remain in the water phase rather than protruding in
the air phase due to their
corresponding small contact angle with this interface.
As shown in VideoS2 in~\cite{EPAPS},
close to the water/air interface the chain propels in
opposite direction.
Although the interfaces in both cases bind the motion of the colloidal
particles and induce a rectification of the rotating colloids,
the local flow
induced by the stick boundary condition characteristic
of the solid wall differs from the one
generated by the slip boundary condition imposed by the
liquid/air interface, leading to an opposite displacement
of the worms. We note that similar effect
has been used in~\cite{Leo11} to reverse the rotation
of {\it E. coli}. We
also find that the average speed of the colloidal worm
at the water/air interface ($\langle {\mathrm v} \rangle = -1.8 \mu m s^{-1}$)
is $25\%$ higher than the
speed close to the solid surface,
$\langle {\mathrm v} \rangle = 1.5\mu m s^{-1}$. \\
In order to explain the hydrodynamic origin
of the observed rectification effect, we propose a model where the
effect of a planar interface on the  motion of  a
solid sphere located at a distance $h$ can
be accounted for, to  lowest order, through an appropriate set of 
hydrodynamic singularities at the  same 
distance $h$ below the position of
the interface. For a liquid/air interface, the image 
reduces to a  particle rotating  in the opposite sense that 
the actuated colloid, while  for a solid surface an additional 
stresslet and source doublet must be added~\cite{Bla74}.\\
A colloid subject to a torque ${\bm T}$ will rotate
and generate a flow field, ${\bf v}$,
around it that will decay asymptotically
as ${\bf v}= -{\bm T}\times {\bm r}/(8\pi \eta r^3)$
where ${\bm r}$ stands for the position vector from the
colloid center. In the presence of a liquid interface, the
induced flow decays as 
${\bf v}= -{\bm T}\times {\bm r}/(8\pi \eta r^3)-{\bm T}^*\times {\bm r}^*/(8\pi \eta r^3)$
where ${\bm r}^*$ is the position vector from the center of the 
image and ${\bm T}^* = - \eta{\bm T}/\eta_g$,
being $\eta_g$ the gas (air) shear viscosity. For a solid interface additional 
terms must be included, as described in~\cite{EPAPS}, and ${\bm T}^* = -{\bm T}$. 
Thus for a liquid/solid
interface the torque acting on the image has the same
magnitude that the one actuating on the deposited particle,
but for a liquid/air interface the image torque
increases by a factor $\eta/\eta_g$~\cite{Pag98}.
Hydrodynamic
rectification emerges from the flow induced by the image singularities
because the flow will  generally have a
component parallel to the interface. As a result, a particle
close to an actuated one will be displaced by such induced flow. We can quantify the  effect
of this displacing flow on an array on $N$ equi-spaced colloids
placed parallel to the wall by computing the velocity
at the center of such an assembly ${\bf v}_{c}$. 
If we consider  that each colloid moves with the total flow
velocity induced by the rest of  rotating colloids
at its position (hence assuming it experiences no
friction force with respect to the local flow), we can arrive at~\cite{EPAPS}:
\begin{eqnarray}
{\bf v}_{c}={\mathrm v}_0 {\bm e}_x+\frac{|{\mathrm v}_0|}{N}\sum_{j=0}^{N}\sum_{i\neq j} \left[1-\xi+\xi\left(\frac{a}{h}\right)^{-2}\right] \times\nonumber \\
\left\{\frac{1-\xi}{[1+\epsilon^2 (i-j)^2]^{3/2}}+ \frac{3 \xi \epsilon^2 (i-j)^2}{[1+\epsilon^2 (i-j)^2]^{5/2}}\right\}{\bm e}_x
\label{eq1}
\end{eqnarray}
which shows that the instantaneous velocity depends
on the worm length, $N$, and the geometrical parameter,
$\epsilon \equiv \delta /2 h$, where $\delta$ is the 
center-to-center separation between consecutive colloids in 
the array, while $\xi=1(0)$ for a solid (liquid) substrate.
Here ${\mathrm v}_0$ stands for the rectifying
velocity of an isolated colloid placed at a distance $h$ above
the liquid/solid interface ${\mathrm v}_0= T_0a^2/(32\pi \eta h^4)$,
while  ${\mathrm v}_0= -T_0/(32\pi \eta_g h^2)$
for a liquid/gas interface. 
Therefore, one can understand both the increase
in the magnitude of the speed of a worm in the presence of a
liquid/gas interface
and the change in direction with respect to the solid/liquid interface. Moreover,
Eq.~(\ref{eq1}) illustrates that the difference between
interfaces enters through the  overall magnitude ${\mathrm v}_0$ and 
that the remaining dependence in the worm size and elevation 
\begin{figure}[t]
\begin{center}
\includegraphics[width=\columnwidth,keepaspectratio]{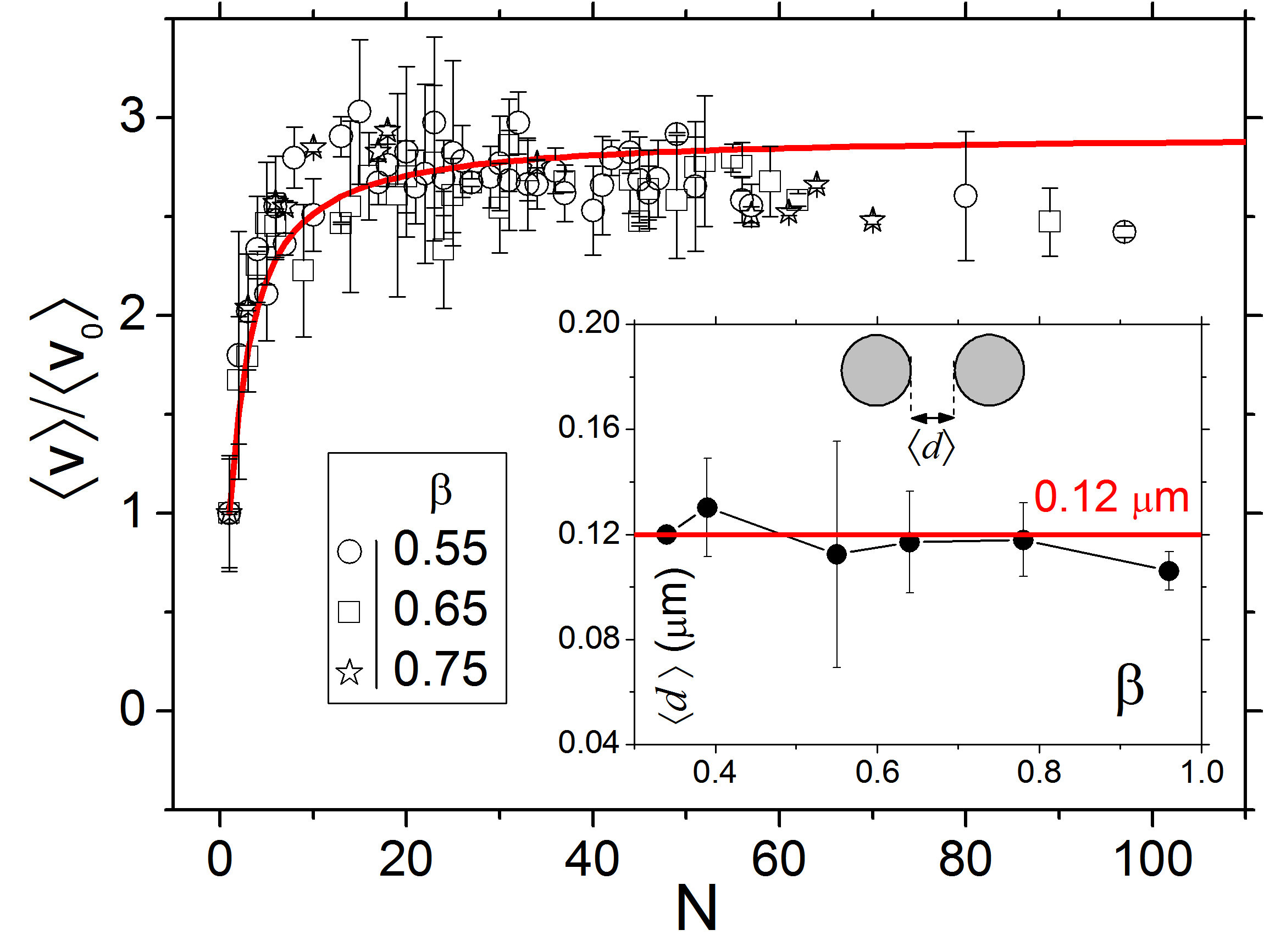}
\caption{(Color online) Normalized average speed $\langle {\mathrm v} \rangle/\langle {\mathrm v}_0 \rangle$
of colloidal worms as a function of
the number of rotors $N$ for different
ellipticity $\beta$ ($\omega=62.8$ rad\, s$^{-1}$, $H_0=2300 A \, m^{-1}$).
Points denote experimental data while the 
continuous line is a fit following Eq.(1).
Inset shows the average surface
distance $\langle d \rangle$
{\it vs} $\beta$.}
\label{fig_3}
\end{center}
\end{figure}
from the interface are purely geometrical.\\
To test the model, we measure the time average 
velocity $\langle {\mathrm v} \rangle$
for a series of
worms characterized by different
number of particles
and at different values of $\beta$. Fig.3 shows the
comparison between the experiment
and theory by fitting the rescaled
velocity $\langle {\mathrm v} \rangle/\langle {\mathrm v}_0 \rangle$
with Eq.(1).
First, the data show that the propulsion
speed is almost independent on $\beta$
at parity of actuating amplitude of the applied field.
Moreover, we use only one adjustable variable as a fitting parameter,
finding $\epsilon=0.92$.
From particle tracking we measure the average distance between two particles $\langle \delta \rangle= \langle d \rangle+2a$, being $\langle d \rangle$ the distance between their surfaces, and confirm that this quantity is indeed independent on $\beta$, and given by $\langle d \rangle =120 nm$, as shown in the inset of Fig.3.
Close contact ($\langle d \rangle =0$) between the particle
is prevented by the electrostatic double layer 
interactions
between the particles
and steric repulsion due to the polymer coating.
The value of $\langle d \rangle$ allows
us to find an elevation
of the worm from the substrate 
$h=1.58 \mu m$ (surface distance from the glass plane
$\sim 180 nm$),
similar to the one found by evanescent light scattering
of individual particles~\cite{Bli05}.
The quantitative agreement between 
experiments and theory,
despite the simplifications used in the model~\cite{EPAPS},
demonstrates that the latter well captures
all the basic features observed,
and provides a clear
description of the physical mechanism
underlying the worm propulsion.\\
A moving worm generates a cooperative flow field
which continuously advects
the colloidal chain in a similar way to 
a hydrodynamic "conveyor-belt"~\cite{Fel12}.
This effect is illustrated in Fig.4,
\begin{figure}[tb]
\begin{center}
\includegraphics[width=\columnwidth,keepaspectratio]{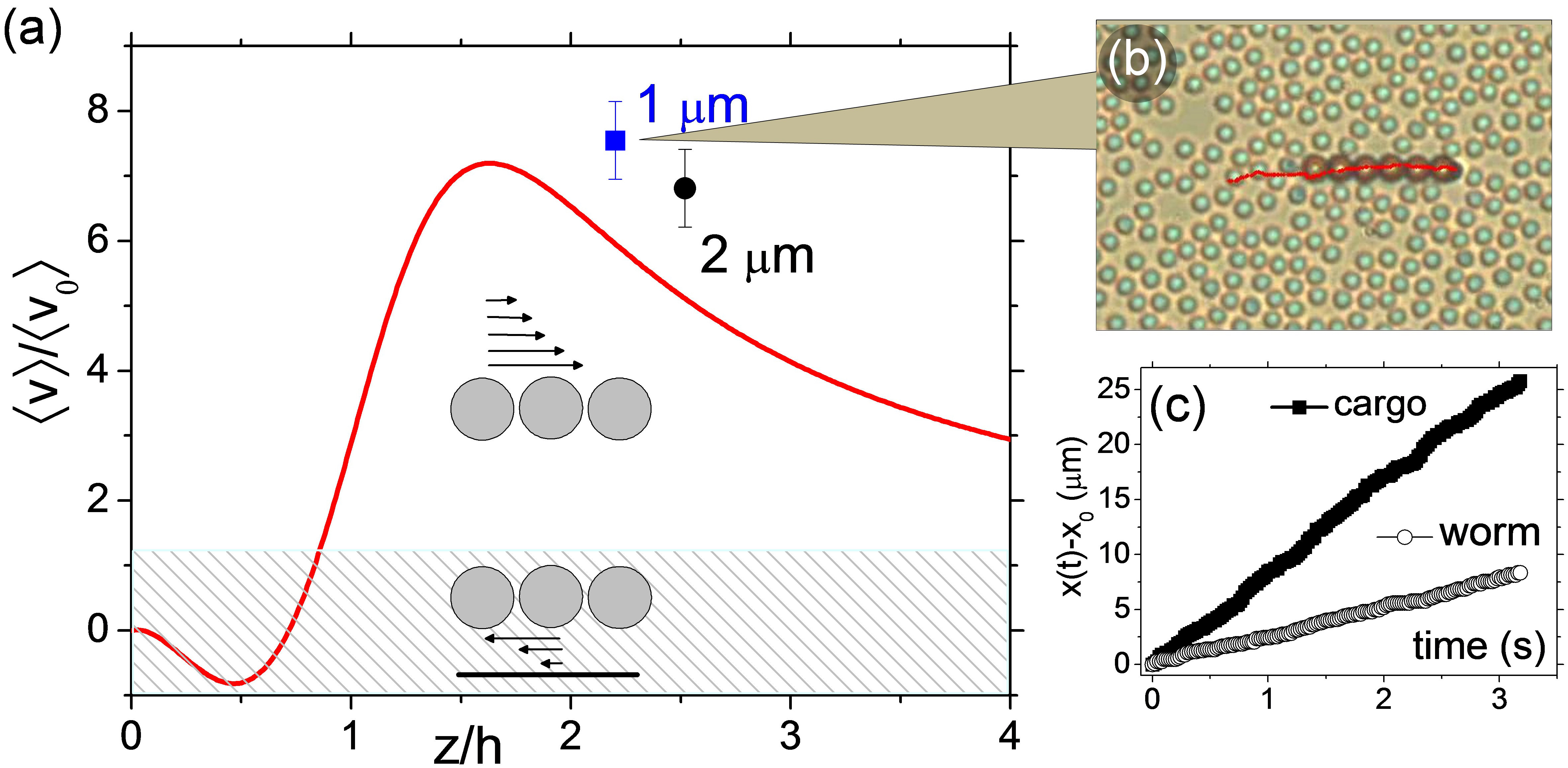}
\caption{(Color online) (a) Normalized average flow speed $\langle {\mathrm v}_f \rangle/\langle {\mathrm v}_0 \rangle$
{\it vs} elevation from the surface $z/h$ and calculated
following the model in~\cite{EPAPS}.
(b) Colloidal worm dragging a $2\mu m$ silica particle (red trajectory),
VideoS3 in~\cite{EPAPS}. (c)
Plots of the distance versus time for the worm (empty circles)
and silica particle (filled squares).}
\label{fig_3}
\end{center}
\end{figure}
where we plot
the normalized average flow velocity $\langle {\mathrm v}_f \rangle/\langle {\mathrm v}_0 \rangle$
produced
by a moving worm as a function of the rescaled elevation
$z/h$ (details of the calculation
are given in~\cite{EPAPS}).
In the small region between the worm
and the bounding plate
the flow velocity is negative,
while it changes sign above
the moving chain.
The flow field can be visualized
by tracking the position of non-magnetic
silica particles used as tracers
and floating
above the plate similar to particle image velocimetry,
Fig.4(b).
Once close to the moving chain,
the tracers are captured by the
hydrodynamic belt and
rapidly dragged
along the worm till
they are expelled at the front, as shown
in VideoS3 in~\cite{EPAPS}.
We confirm this mechanism by measuring the average speed
of two different colloidal "cargos"
having sizes $1\mu m$ and $2\mu m$
shuffled by the conveyor belt at an
average speed of $8.3 \mu m s^{-1}$
and $7.4 \mu m s^{-1}$ resp. (here ${\mathrm v}_0=1.1 \, \mu m s^{-1}$).\\
In conclusion, we have demonstrated
and theoretically supported
a concept of directed propulsion at
low Reynolds number
based on the use of magnetically assembled
colloidal rotors. Propulsion arises due to the 
non-linear cooperative rectification of flows
generated by the spinning particles close to the bounding wall.
The hydrodynamic flow generated by our
magnetic prototype can be used as an efficient
mechanism to transport biological or colloidal cargos
entrapped and translated by the conveyor belt.
On the other hand, trapping the leading rotor
of the worm by optical tweezers could convert
the magnetic propeller into a fluid pump
which can be readily employed in micro- and nanofluidc systems.

\begin{acknowledgments}
F.M.P., A.O.A. and P.T. acknowledge support from the
ERC starting Grant "DynaMO".
P.T. acknowledges support from
the "Ramon y Cajal" Program No.RYC-2011-07605, 
from MINECO (FIS2013-41144-P) and DURSI (2014SGR878).
I.P. acknowledges support from MINECO (Spain),
Project FIS2011-22603, DURSI Project 2014SGR-922,
and Generalitat de Catalunya under Program "ICREA  Acad\`emia".
\end{acknowledgments}
\bibliography{biblio}
\end{document}